\documentclass{article}[9pt]
\usepackage{spconf,amsmath,graphicx}
\usepackage{amssymb}
\usepackage{amsthm,bm}
\usepackage{optidef,url}
\newcommand{\be}{\mathbf{e}}
\newcommand{\bx}{\mathbf{x}}

\newcommand{\by}{\mathbf{y}}

\newcommand{\bg}{\mathbf{g}}
\newcommand{\bY}{\mathbf{Y}}

\newcommand{\bR}{\mathbf{R}}
\newcommand{\bs}{\mathbf{s}}

\newcommand{\bS}{\mathbf{S}}

\newcommand{\bH}{\mathbf{H}}

\newcommand{\bI}{\mathbf{I}}

\newcommand{\Ball}{\mathcal{B}}

\newcommand{\Pcal}{\mathcal{P}}

\newcommand{\Scal}{\mathcal{S}}
\newcommand{\Pex}{\Pcal_{\text{ex}}}
\newcommand{\bD}{\mathbf{D}}
\newcommand{\bPi}{\bm{\Pi}}

\newtheorem{lemma}{Lemma}
\newtheorem{definition}{Definition}
\title{AN INFORMATION MAXIMIZATION BASED BLIND SOURCE SEPARATION APPROACH FOR DEPENDENT AND INDEPENDENT SOURCES}
\name{Alper T. Erdogan\thanks{This work is partially supported by a funding provided by the
KUIS AI Lab.}}
\address{EE Department and KUIS AI Lab, Koc University}
\begin{document}
%\ninept
%
\maketitle
\begin{abstract}
We introduce a new information maximization (infomax) approach for the blind source separation problem. The proposed framework provides an information-theoretic perspective for determinant maximization-based structured matrix factorization methods such as nonnegative and polytopic matrix factorization. For this purpose, we use an alternative joint entropy measure based on the log-determinant of covariance, which we refer to as log-determinant (LD) entropy. The corresponding (LD) mutual information between two vectors reflects a level of their correlation. We pose the infomax BSS criterion as the maximization of the LD-mutual information between the input and output of the separator under the constraint that the output vectors lie in a presumed domain set.  In contrast to the ICA infomax approach, the proposed information maximization approach can separate both dependent and independent sources. Furthermore, we can provide a finite sample guarantee for the perfect separation condition in the noiseless case.
\end{abstract}
\section{Introduction}
\label{sec:intro}
Blind source separation (BSS)  has long been a central problem of interest in both the signal processing and machine learning communities \cite{comon2010handbook,comon1994independent,hyvarinen2001independent}. Solution approaches proposed for this problem exploit side information regarding the data model to extract latent components from their mixtures.

Independent component analysis (ICA) has been a key framework that utilizes the assumption about the mutual independence of sources. Among several alternative ICA algorithmic approaches, the infomax principle provided an information-theoretic perspective to the BSS problem \cite{bell1995information}. It aims to maximize the information transfer from input to output while constraining the output to satisfy the mutual independence assumption. 

As an alternative to ICA, we can list deterministic approaches that utilize particular domain structures of the sources. For example, nonnegative matrix factorization (NMF) \cite{paatero1994positive,lee1999learning,fu2016robust,fu2019nonnegative} exploits the assumption that the sources are located in the nonnegative orthant. Antisparse bounded component analysis (BCA) algorithms make use of the sources' $\ell_\infty$-norm ball membership feature  \cite{cruces2010bounded,erdogan2013class,inan2014convolutive}. Sparse component analysis (SCA) algorithms can be based on the assumption that the sources are from the $\ell_1$-norm ball\cite{elad2010sparse,georgiev2005sparse,babatas2018algorithmic}. Recently, polytopic matrix factorization (PMF) has been introduced as a new framework for solving inverse linear problems based on the knowledge that source vectors are drawn from polytopes \cite{tatli:21icassp,tatli:21tsp}.  A common separation criterion employed by all these deterministic frameworks is the maximization of the determinant of the sample correlation matrix of source estimates.

This article introduces a new information maximization perspective for  BSS approaches based on determinant maximization. The proposed  framework differs from the ICA-infomax approach from \cite{bell1995information} in two major ways: 
\begin{itemize}
\item[(i)] it does not necessarily assume that sources are mutually independent, and %therefore, it applies to both dependent and independent component scenarios,
\item[(ii)] instead of  Shannon-mutual information, it uses an alternative, the so-called log-determinant (LD)-mutual information measure (see Section \ref{sec:LD}).
\end{itemize}
Regarding item (ii) above, for a vector with the covariance matrix $\bR_\bx$, we use a joint entropy measure (LD-entropy) based on the log-determinant of the covariance matrix, which disregards the higher order statistics about the random vector. As we will show in Section \ref{sec:LD}, the corresponding mutual information reflects correlation rather than dependence between its arguments. Furthermore, it is based on a linear inference model with the minimum mean square error (MMSE) criterion. The LD-entropy measure has recently been proposed as an estimator of Shannon entropy for the information bottleneck analysis of deep neural networks \cite{zhouyin:21}. However, we do not use it as an approximation tool; instead, we use it as a standalone measure relevant to the linear inference setting of the BSS problem. The resulting optimization objective function resembles but is different from those used in determinant maximization-based matrix factorization approaches \cite{fu2016robust,fu2019nonnegative,tatli:21tsp}.

The following is the article's organization: Section \ref{sec:BSSsetup} provides the BSS setup used throughout the article. Although we use PMF as a reference approach, it can be adapted to structured matrix factorization approaches such as NMF. Section \ref{sec:pmf} summarizes results relevant to BSS from PMF framework. In Section \ref{sec:LD}, we introduce the LD-entropy and LD-mutual information measures
and their deterministic counterparts. We define the LD-infomax criterion for BSS in Section \ref{sec:ldinfomax} and provide a corresponding algorithm in Section \ref{sec:algorithm}. Finally, in Section \ref{sec:example}, we provide an example that illustrates the LD-infomax algorithm's dependent source separation capability.
\section{Source Separation Setup}
\label{sec:BSSsetup}

We assume the following generative data model for  observations throughout the article:
 
 \underline{\it Sources:} There are $r$ real valued sources with $N$ samples each, which form
    $\mathcal{S}_g=\{\bs_g(1), \ldots, \bs_g(N)\} \subset \mathbb{R}^r$. 
    We also define the matrix for the generative source samples as
    \begin{eqnarray*}
    \bS_g=\left[\begin{array}{cccc} \bs_g(1) & \bs_g(2) & \ldots & \bs_g(N) \end{array}\right]\in \mathbb{R}^{r \times N}.
    \end{eqnarray*}
    We further assume that $\Scal_g$ is a subset of a polytope $\Pcal$. The choice of $\Pcal$ reflects the assumptions on sources and their mutual relationships \cite{tatli:21icassp,tatli:21tsp}. For example, defining $\Ball_p=\{\bx \vert \|\bx\|_p\le 1\}$ as the $\ell_p$-norm ball:
    \begin{itemize}
        \item $\Pcal=\Ball_1$ ($\Pcal=\Ball_\infty$) corresponds to sparse  (anti-sparse) sources, and
        \item $\Pcal=\Ball_{1,+}=\Ball_{1}\cap \mathbb{R}_+^r$ ($\Pcal=\Ball_{\infty,+}=\Ball_\infty\cap\mathbb{R}_+^r$) is for sparse (anti-sparse) nonnegative sources.
    \end{itemize}
    In addition to these polytopes corresponding to globally defined source attributes, we can also define polytopes with heterogeneous source properties, such as the example provided in \cite{tatli:21tsp}:
    \begin{eqnarray*}
\def\arraystretch{1.6}
\hspace*{0.04in}\Pex=\left\{\mathbf{s}\in \mathbb{R}^3\ \middle\vert \begin{array}{l}   s_1,s_2\in[-1,1],s_3\in[0,1],\\ \left\|\left[\begin{array}{c} s_1 \\ s_2 \end{array}\right]\right\|_1\le 1,\left\|\left[\begin{array}{c} s_2 \\ s_3 \end{array}\right]\right\|_1\le 1 \end{array}\right\},
\end{eqnarray*}
where $s_1,s_2$ are signed, $s_3$ is nonnegative and the sparsity is defined between $s_1,s_2$ and $s_2, s_3$. The PMF framework offers infinite choices for $\Pcal$, which provides a high degree of freedom in generating diverse sets of source features and relationships. For a polytope to lead to a "solvable" BSS problem, it needs to be "identifiable" in the sense that  only linear transformations that map $\Pcal$ to itself are permutation or diagonal scaling operations \cite{tatli:21tsp}.

To guarantee the identifiability of the original sources, the PMF framework in \cite{tatli:21tsp} proposed a sufficiently scattered condition for the source samples in $\Scal_g$ based on the maximum volume inscribed ellipsoid (MVIE) of $\Pcal$ \cite{boyd2004convex}. This condition requires that the source samples' convex hull, $\text{conv}(\Scal_g)$, forms a better approximation of the polytope than its MVIE, as formalized by the following definition: 
\begin{definition}\label{def:suffscat}{\it Sufficiently Scattered Set of $\Pcal$}: 
 $\Scal_g$ is called a sufficiently scattered set of  $\mathcal{P}$ if  
\begin{itemize}
\item[]{\it (PMF.SS.i)} $\mathcal{P} \supseteq \text{conv}(\Scal_g)\supset \mathcal{E}_\mathcal{P}$, and
\item[]{\it (PMF.SS.ii)} $\text{conv}(\bS)^{*,\bg_\mathcal{P}} \cap \text{bd}(\mathcal{E}_\mathcal{P}^{*,\bg_\mathcal{P}})=\text{ext}(\mathcal{P}^{*,\bg_\mathcal{P}})$,
\end{itemize}
where $\mathcal{E}_\mathcal{P}$ is the MVIE of $\mathcal{P}$, centered at $\bg_\mathcal{P}$, $\text{bd}(\cdot)$ and $\text{ext}(\cdot)$ stand for boundary and vertices respectively.
\end{definition}
The notation $C^{*,\bg}$ in the (PMF.SS.ii) above stands for the polar of the set $C\subset \mathbb{R}^r$ relative to a point $\bg$ as defined by:
\begin{eqnarray*}
C^{*,\mathbf{g}}=\{ \bx \in \mathbb{R}^r | \langle \bx,\by-\mathbf{g} \rangle \leq 1 \ \forall \by\in C\}.
\end{eqnarray*}
In Definition \ref{def:suffscat}, (PMF.SS.i) ensures that $\text{conv}(\Scal_g)$ contains $\mathcal{E}_\mathcal{P}$ while (PMF.SS.ii) restricts how tight the boundary of $\text{conv}(\Scal_g)$ encloses $\mathcal{E}_\mathcal{P}$ by restricting its tangency points.

\underline{\it Mixing:} We assume that sources are mixed through linear mapping with matrix $\bH_g\in \mathbb{R}^{M\times r}$, where we consider the (over)determined case, i.e., $M\ge r$. In the noiseless case, we can write the observations, i.e., the mixture vectors, as:
\begin{eqnarray}
\by(k)=\bH_g\bs_g(k), \hspace{0.1in} k \in \{1, \ldots, N \}.
\end{eqnarray}
We can write the corresponding matrix of mixtures as
\begin{eqnarray}
\bY&=&\left[\begin{array}{cccc} \by(1) & \by(2) & \ldots & \by(N)\end{array}\right]=\bH_g \bS_g. \nonumber 
\end{eqnarray}
%\end{itemize}
In BSS, the goal is to estimate $\bS_g$ from  $\bY$ when $\bH_g$ is not known and no training information is available. 

\section{PMF as a  BSS Approach}
\label{sec:pmf}
In this section, we revisit the PMF approach introduced in \cite{tatli:21icassp,tatli:21tsp} which can be utilized to solve the BSS problem in Section \ref{sec:BSSsetup}. The PMF approach exploits the information that the original source samples are  sufficiently scattered inside a polytope $\Pcal$. The corresponding criterion is defined in terms of the following optimization problem:
\begin{maxi!}[l]<b>
{\bH\in \mathbb{R}^{M \times r},\bS\in \mathbb{R}^{r\times N}}
{\det(\bS\bS^T)\label{eq:detmaxobjective}}{\label{eq:detmaxoptimization}}{}
\addConstraint{\bY=\bH \bS}{\label{eq:detmaxconstr1}}{}
\addConstraint{\bS_{:,j} \in \mathcal{P}}{,\quad}{j= 1, \ldots, N.\label{eq:detmaxconstr2}}{ }
\end{maxi!}
The objective function in (\ref{eq:detmaxobjective}) is the scaled sample correlation matrix, which is a measure of the spread of the columns of $\bS$ inside $\Pcal$. According to \cite{tatli:21icassp,tatli:21tsp}, if
\begin{itemize}
    \item[I.] $\Pcal$ is an "identifiable" polytope, as defined in \cite{tatli:21tsp}, and %Definition \ref{def:idpoly}, and
    \item[II.] the columns of $\bS_g$ are a "sufficiently scattered set" in $\Pcal$ as defined in Definition \ref{def:suffscat},
\end{itemize}
then any $\bS_*$ of (\ref{eq:detmaxoptimization}) can be written as:
 $\bS_*=\bPi\bD\bS_g$ 
where $\bD$ is a diagonal matrix with $\pm 1$ diagonal entries, corresponding to sign ambiguity, and $\bPi$ is a permutation matrix, corresponding to permutation ambiguity. 
We can modify the PMF optimization problem in (\ref{eq:detmaxoptimization}) by replacing the objective with the log-determinant of the sample covariance matrix:
$\hat{\bR}_\bs=\frac{1}{N}{\bS}{\bS}^T-\frac{1}{N^2}{\bS}\mathbf{1}\mathbf{1}^T{\bS}^T$, 
as
\begin{maxi!}[l]<b>
{\bH\in \mathbb{R}^{M \times r},\bS\in \mathbb{R}^{r\times N}}
{\log\det(\hat{\bR}_\bs)\label{eq:detmax2objective}}{\label{eq:detmax2optimization}}{}
\addConstraint{\bY=\bH \bS}{\label{eq:detmax2constr1}}{}
\addConstraint{\bS_{:,j} \in \mathcal{P}}{,\quad}{j= 1, \ldots, N.\label{eq:detmax2constr2}}{ }
\end{maxi!}
In the case of noisy mixtures, we can remove the equality constraint in (\ref{eq:detmax2constr1}) and replace the objective function with \cite{fu2016robust,tatli:21tsp}:
\begin{eqnarray}
\label{eq:noisypmf}
J(\bS)=\frac{1}{2}\log\det(\hat{\bR}_\bs)-\gamma\|\bY-\bH\bS\|_F^2.
\end{eqnarray}

\section{Log-Determinant Entropy and Information Measure}
\label{sec:LD}
We define the log-determinant (LD)-entropy of a random vector $\bx\in\mathbb{R}^r$ with covariance $\bR_\bx$ as \cite{zhouyin:21}
\begin{eqnarray}
\label{eq:HLD}
H_{LD}^{(\epsilon)}(\bx)=\frac{1}{2}\log\det(\bR_\bx+\epsilon \bI)+\frac{r}{2}\log(2\pi e),
\end{eqnarray}
where $\epsilon>0$ is a small value to ensure $\log\det$ term is lower bounded by a finite value. 
Note that $H_{LD}^{(0)}(\bx)$ is equal to the expression for Shannon differential entropy for a Gaussian vector, which constitutes an upper bound on the Shannon differential entropy for all random vectors with covariance $\bR_\bx$.

The joint LD-entropy of  $\bx\in\mathbb{R}^r$ and $\by\in\mathbb{R}^q$ can be written as:
\begin{eqnarray}
&&\hspace*{-0.2in}H_{LD}^{(\epsilon)}(\bx,\by)=\frac{1}{2}\log\det(\bR_{\scriptsize\left[\begin{array}{c}\bx\\ \by\end{array}\right]}+\epsilon \bI)+\frac{n+m}{2}\log(2\pi e), \nonumber \\
&&\hspace*{-0.3in}=\frac{1}{2}\log\det\left(\left[\begin{array}{cc} \bR_\bx+\epsilon \bI & \bR_{\bx\by} \\ \bR_{\bx\by}^T & \bR_{\by}+ \epsilon \bI \end{array}\right]\right)+\frac{r+q}{2}\log(2\pi e), \nonumber \\
&&\hspace*{-0in}=\frac{1}{2}\log\det(\bR_\bx+\epsilon \bI)+\frac{r}{2}\log(2\pi e)\nonumber \\ &&\hspace*{-0in}+\frac{1}{2}\log\det(\bR_\be+\epsilon \bI)+\frac{q}{2}\log(2\pi e) \nonumber \\
&&\hspace*{-0in}=H_{LD}^{(\epsilon)}(\bx)+H_{LD}^{(\epsilon)}(\by \vert_L \bx),\nonumber
\end{eqnarray}
where $\bR_\bx$,$\bR_\by$ are the (auto)covariance matrices of $\bx$ and $\by$ vectors respectively,  $\bR_{\bx\by}$ is the cross covariance matrix between $\bx$ and $\by$, and $\bR_\be=\bR_\by-\bR_{\bx\by}^T(\bR_\bx+\epsilon \bI)^{-1}\bR_{\bx\by}$.
Note that $H_{LD}^{(\epsilon)}(\by \vert_L \bx)=\frac{1}{2}\log\det(\bR_\be+\epsilon \bI)+\frac{q}{2}\log(2\pi e)$ is not equal to the LD-entropy of $\by \vert \bx$ which is expressed by $H_{LD}(\by \vert \bx)=\frac{1}{2}\log\det(\bR_{\by|\bx})+\frac{q}{2}\log(2\pi e)$.  The notation $\by \vert_L \bx$ signifies this difference. 

If we consider the limit $\epsilon\rightarrow 0$, the expression  $H_{LD}^{(\epsilon)}(\by \vert_L \bx)$ corresponds to the log-determinant of the error covariance of the best affine (MMSE) of  $\by$ from $\bx$. Therefore, $H_{LD}^{(\epsilon)}(\by \vert_L \bx)$ can be considered a measure of the uncertainty remaining in $\by$ after linearly estimating it from $\bx$ with respect to the MMSE criterion, for sufficiently small $\epsilon$.

Based on the LD-entropy definition in (\ref{eq:HLD}), we can provide the corresponding LD-mutual information measure as:
\begin{eqnarray}
I_{LD}^{(\epsilon)}(\bx,\by)&=&H_{LD}^{(\epsilon)}(\by)-H_{LD}^{(\epsilon)}(\by \vert_L \bx).
\end{eqnarray}
We can show also show that: 
\begin{eqnarray}
I_{LD}^{(\epsilon)}(\bx,\by)&=&H_{LD}^{(\epsilon)}(\bx)-H_{LD}^{(\epsilon)}(\bx \vert_L \by)%\\
%                      &=& H^{(\epsilon)}(\bx)+H^{(\epsilon)}(\by)-H^{(\epsilon)}(\bx,\by).
\end{eqnarray}
For LD-mutual information, the following property holds:
\begin{lemma}
For non-degenerate random vectors $\bx$ and $\by$ with $\bR_\bx\succ \mathbf{0}$ and $\bR_\by \succ\mathbf{0}$ respectively,  $I^{(\epsilon)}_{LD}(\bx,\by)\ge 0$ with equality iff they are mutually uncorrelated, i.e. $\bR_{\bx\by}=\mathbf{0}$.
\end{lemma}
In summary, LD-entropy and LD-mutual information act as relevant uncertainty and correlative information measures, respectively, when the inference is based on the linear MMSE estimation, for $\epsilon$ sufficiently small.  

\subsection{Deterministic Case}
For a deterministic setting such as the BSS setup in Section \ref{sec:BSSsetup}, we can work with the deterministic counterparts of the LD-entropy and LD-mutual information provided above. For example, for a set of source samples $\{\bs(1), \bs(2), \ldots, \bs(N)\}$ with the sample covariance $\hat{\bR}_\bs$, the corresponding deterministic LD-entropy can be written as:
\begin{eqnarray}
\label{eq:detHLD}
\hat{H}_{LD}^{(\epsilon)}(\bS)=\frac{1}{2}\log\det(\hat{\bR}_\bs+\epsilon \bI)+\frac{r}{2}\log(2\pi e).
\end{eqnarray}
We can, therefore, consider that the PMF optimization objective in (\ref{eq:detmax2objective}) is essentially equivalent to $\hat{H}_{LD}^{(\epsilon)}(\bs)$.
It is interesting  that  for a sufficiently scattered $\Scal_g$ of $\Pcal$, no linear transformation that maps $\Scal_g$ into $\Pcal$  generates a set with larger deterministic LD-entropy, which is implied by the identifiability condition of the PMF setting in (\ref{eq:detmax2optimization}) \cite{tatli:21tsp}. Therefore, one can consider a sufficiently scattered set $\Scal_g$ in $\Pcal$ as a maximal deterministic LD-entropy set of $\Pcal$ in this sense.

The deterministic LD-mutual information for $\bs$ and $\by$ is expressed as:
\begin{eqnarray}
\label{eq:detldmi}
\hat{I}_{LD}^{(\epsilon)}(\bY,\bS)&=&\frac{1}{2}\log\det(\hat{\bR}_{\bs}+\epsilon\bI)\nonumber \\&&\hspace*{-0.8in}-\frac{1}{2}\log\det(\hat{\bR}_\bs-\hat{\bR}_{\bs\by}(\epsilon \bI+\hat{\bR}_\by)^{-1}\hat{\bR}_{\bs\by}^T+\epsilon \bI),
\end{eqnarray}
which is written in terms of the sample covariance matrices
%\begin{eqnarray*}
$\hat{\bR}_\by=\frac{1}{N}{\bY}{\bY}^T-\frac{1}{N^2}{\bY}\mathbf{1}\mathbf{1}^T{\bY}^T$, 
%\end{eqnarray*}
and
%\begin{eqnarray*}
$\hat{\bR}_{\bs\by}=\frac{1}{N}{\bS}{\bY}^T-\frac{1}{N^2}{\bS}\mathbf{1}\mathbf{1}^T{\bY}^T$.
%\end{eqnarray*}

\section{LD-Infomax as BSS Criterion}
\label{sec:ldinfomax}
To obtain a solution to the BSS setting in Section \ref{sec:BSSsetup}, based on the LD-mutual information measure defined in Section \ref{sec:LD}, we propose the use of the following optimization problem:
\begin{maxi!}[l]<b>
{\bS\in \mathbb{R}^{r\times N}}
{\hat{I}_{LD}^{(\epsilon)}(\bY,\bS)\label{eq:infomaxobjective}}{\label{eq:infomaxoptimization}}{}
\addConstraint{\bS_{:,j} \in \mathcal{P}}{,\quad}{j= 1, \ldots, N.\label{eq:infomaxconstr2}}{ }
\end{maxi!}
The proposed optimization aims to maximize the LD-mutual information between the output samples $\{\bs(1), \ldots, \bs(N)\}$ and input mixture samples $\{\by(1), \ldots \by(N)\}$ while the output samples are constrained to lie in $\Pcal$. This approach  exploits the assumption that the original source samples form a maximal LD-entropy set of $\Pcal$. The objective function in (\ref{eq:detldmi}) resembles the form in (\ref{eq:noisypmf}), of which the second term is replaced with  $\frac{1}{2}\log\det(\hat{\bR}_\be+\epsilon \bI)$, where $\hat{\bR}_\be=\hat{\bR}_\bs-\hat{\bR}_{\bs\by}(\epsilon \bI+\hat{\bR}_\by)^{-1}\hat{\bR}_{\bs\by}^T$.

In the noiseless BSS setting, when $\epsilon$ is selected to be sufficiently small, the second term in (\ref{eq:detldmi}) is maximized when the relation between the input and output is linear. Furthermore, the first term in (\ref{eq:detldmi}) approaches the objective function in (\ref{eq:detmax2objective}). Therefore, the optimization problem in (\ref{eq:infomaxoptimization}) becomes equivalent to the PMF optimization in (\ref{eq:detmax2optimization}). If  $\Scal_g$ is sufficiently scattered, i.e., a maximal LD-entropy set of $\Pcal$, then it is possible to identify the original sources, with some permutation and sign ambiguity, from a finite number of samples.

\section{Algorithm}
\label{sec:algorithm}
As a simple approach, we consider a projected gradient search  algorithm for the optimization problem in (\ref{eq:infomaxoptimization}). The gradient of the objective function in (\ref{eq:infomaxobjective}) is given by:
\begin{eqnarray}
&&\nabla_\bS \hat{I}_{LD}^{(\epsilon)}(\bY,\bS)=\frac{1}{N}(\hat{\bR}_\bs+\epsilon\bI)^{-1}\bS(\bI-\frac{1}{N}\mathbf{11}^T)\nonumber\\
&&\hspace*{-0.3in} -\frac{1}{N}(\hat{\bR}_\be+\epsilon\bI)^{-1}(\bS-\hat{\bR}_{\bs\by}(\hat{\bR}_\by+\epsilon\bI)^{-1}\bY)(\bI-\frac{1}{N}\mathbf{11}^T).\nonumber
\end{eqnarray}
We can express the corresponding  algorithm update rule as:
\begin{eqnarray}
\bS^{(k+1)}=P_\Pcal\left(\bS^{(k)}+\mu^{(k)}\nabla_\bS \hat{I}^{(\epsilon)}(\bY,\bS^{(k)})\right),
\end{eqnarray}
where $P_\Pcal(\cdot)$ is the operator for projection onto the polytope $\Pcal$, and $\mu^{(k)}$ is the learning rate at iteration $k$.

\section{Numerical Example}
\label{sec:example}
%\subsection{Correlated Nonnegative Sources}
We consider a BSS scenario with $r=5$ sources and $M=8$ mixtures. The sources are nonnegative copula-t random variables with $4$-degrees of freedom with a Toeplitz generator correlation matrix of which the first row is given by $\left[\begin{array}{ccccc} 1 & \rho & \rho & \rho & \rho \end{array}\right]$. Here, $\rho$ is the parameter that controls the correlation between  sources. Therefore, the columns of $\bS_g$ are correlated samples in $\Ball_{1,+}$. The mixing matrix $\bH$ is generated as a random $8\times 5$ matrix with i.i.d. zero mean and unity variance Gaussian entries. The mixtures are corrupted by i.i.d. Gaussian noise. We used $N=10000$ samples at each realization and chose $\epsilon=1\times10^{-5}$.

In the simulations, at each iteration, we calculated the mean square error $\text{MSE}^{(k)}=\frac{1}{N}\|\bS^{(k)}-\bD^{(k)}\bPi^{(k)}\bS_g\|_F^2$, where $\bD^{(k)}$ and $\bPi^{(k)}$ are diagonal and permutation matrices  that minimize the error. We evaluated the signal-to-interference+noise-power-ratio (SINR) by calculating the ratio of the average source power to $\text{MSE}^{(k)}$.
\begin{figure}[h]
	\centering
	\includegraphics[width=0.65\linewidth,trim={0 0 0 1.5cm},clip]{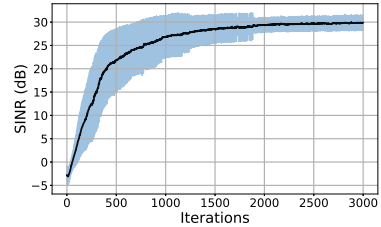}
	\caption{Convergence of the LD-infomax algorithm.}
	\label{fig:convergence}
\end{figure}
We used the step size rule with $\mu^{(k)}=\frac{200}{\sqrt{k+1}}$ for the LD-Infomax algorithm in Section \ref{sec:algorithm}. For a mixture SNR of $30$dB and $\rho=0.5$, Figure \ref{fig:convergence} shows mean SINR convergence curve, for $100$ trials, with standard deviation envelope as a function of iterations. Although the iteration complexity of this batch algorithm is relatively high, it exhibits similar smooth convergence behavior for all the polytope examples in Section \ref{sec:BSSsetup}. We consider lower complexity implementations of  LD-infomax  as future work.

\begin{figure}[h]
	\centering
	\includegraphics[width=0.75\linewidth,trim={0 0 0 1.5cm},clip]{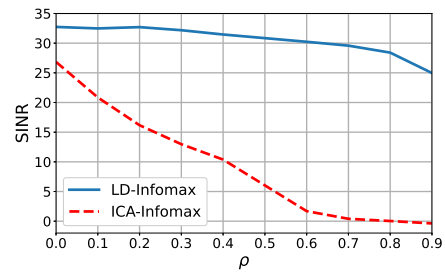}
	\caption{SINR performance comparison of LD and ICA Infomax algorithms.}
	\label{fig:comparison}
\end{figure}

We next compared the proposed LD-infomax algorithm with the ICA-infomax algorithm (the implementation in Python MNE toolbox \cite{GramfortEtAl2013a}) for different correlation parameter $\rho$ values. Since the sources are correlated uniform random variables, we selected the number of sub-Gaussian parameters to be $r=5$. The SINR vs. $\rho$ curves for both algorithms in Figure \ref{fig:comparison} demonstrate that the ICA-infomax algorithm is prone to source correlations, whereas the LD-infomax algorithm is  more immune. In fact, the LD-infomax algorithm  can  successfully separate correlated natural images from their mixtures\footnote{We had to skip an illustrating example due to space limitations.}. We can attribute the slight degradation in the LD-infomax algorithm's performance to the increasing probability of sufficient scattering condition violation with the increasing correlation level.

% Below is an example of how to insert images. Delete the ``\vspace'' line,
% uncomment the preceding line ``\centerline...'' and replace ``imageX.ps''
% with a suitable PostScript file name.
% -------------------------------------------------------------------------
%\begin{figure}[htb]

%\begin{minipage}[b]{1.0\linewidth}
%  \centering
%  \centerline{\includegraphics[width=8.5cm]{image1}}
%%  \vspace{2.0cm}
%  \centerline{(a) Result 1}\medskip
%\end{minipage}
%
%\begin{minipage}[b]{.48\linewidth}
%  \centering
%  \centerline{\includegraphics[width=4.0cm]{image3}}
%%  \vspace{1.5cm}
%  \centerline{(b) Results 3}\medskip
%\end{minipage}
%\hfill
%\begin{minipage}[b]{0.48\linewidth}
%  \centering
%  \centerline{\includegraphics[width=4.0cm]{image4}}
%%  \vspace{1.5cm}
%  \centerline{(c) Result 4}\medskip
%\end{minipage}
%
%\caption{Example of placing a figure with experimental results.}
%\label{fig:res}
%
%\end{figure}

%\vfill
%\pagebreak

%\vfill
%\pagebreak

\newpage

\bibliographystyle{IEEEbib}
\bibliography{ldentropy}

\end{document}